\documentstyle[12pt]{article}
\def\a{\alpha}\def\b{\beta}\def\eps{\epsilon}

\def\be{\begin{equation}}\def\ee{\end{equation}}

\newcommand{\p}[1]{(\ref{#1})}
\begin{document}
\renewcommand{\thefootnote}{\arabic{footnote}}
\begin{flushright}
DFPD 99/TH/53\\
hep-th/9912076
\end{flushright}

\vspace{1truecm}
\begin{center}
{\large\bf Branes in Super--AdS Backgrounds and Superconformal Theories}
\footnote{Talk given by D.S. at the International Workshop
``Supersymmetry and Quantum Symmetries'', JINR, Dubna, Russia, 
July 26--31, 1999}

\bigskip
Paolo Pasti${}^*$, Dmitri~Sorokin ${}^*{}^{\dagger}$\footnote{
On leave from Kharkov Institute of
Physics and Technology, Kharkov, 310108, Ukraine.} 
and Mario Tonin$^*$

\bigskip
${}^*$ Universit\`a Degli Studi Di Padova,
Dipartimento Di Fisica ``Galileo Galilei''\\
ed INFN, Sezione Di Padova
Via F. Marzolo, 8, 35131 Padova, Italia

\bigskip
${}^\dagger$ Humboldt-Universit\"at zu Berlin,\\
Institut f\"ur Physik
Invalidenstrasse 110, D-10115 Berlin, Germany

\bigskip
{\bf Abstract}
\end{center}
We demonstrate how actions for interacting superconformal field theories
in (p+1)-dimensions arise as a result of gauge fixing worldvolume
diffeomorphisms and fermionic $\kappa$-symmetry in actions for
super-p-branes propagating in superbackgrounds of $AdS_{p+2}\times S^{D-p-2}$ 
geometry.
The method of nonlinear realizations and coset spaces is used for getting
an explicit form of supervielbeins and superconnections of the
$AdS\times S$ superbackgrounds, which are required for the construction of the
superconformal theories.
Subtleties of consistent gauge fixing worldvolume symmetries of the branes
are discussed.

\bigskip
During the last two years there have 
been a great interest and an intensive 
development of the Maldacena conjecture \cite{ads} which suggests 
that supergravity 
theories (or, more generally, superstring theories and M--theory) 
on spacetimes 
with a geometry of anti--de--Sitter times a compact manifold (for instance, a 
sphere $S$) are in a certain sense equivalent (or dual) to superconformal 
theories living on a boundary of the AdS space. 
%This implies that a theory in the bulk of
%$AdS\times S$ can be expressed in terms of a conformal theory on the AdS 
%boundary ${\cal M}$, and vice versa. 
More precise statement is that the 
generating functional of correlation functions of observables of 
the conformal 
field theory
on the boundary is equal to the partition function of supergravity (or string)
theory in the bulk \cite{ads}. In a classical approximation this reads
\begin{equation}\label{1}
\big <\exp\int_{\cal M}\Phi_0{\cal 
O}\big>_{CFT}=\exp\left(-S_{AdS}(\Phi)\right),
\end{equation}
where $S_{AdS}(\Phi)$ is the action for the fields $\Phi$ of the bulk theory,
and $\Phi_0$ are boundary values of $\Phi$, 
which are considered as sources for 
conformal fields (operators) ${\cal O}$ of the conformal theory on ${\cal M}$.

%%%%%%%%%%%%%%%%%%%
%For instance, IIB superstring theory in $AdS_5\times S^5$ in a certain limit
%of parameters is dual to 
%${\cal N}=4$ $SU(N)$ Super--Yang--Mills theory on the $D=4$ boundary 
%of $AdS_5$. Parameters of two theories are related as follows
%\begin{equation}\label{para}
%R^4_{AdS}=l_s^4g_sN=l_s^4g^2_{SYM}N, \quad g^2_{SYM}N=\lambda, \quad 
%g_s={\lambda\over N},
%\end{equation}
%where $R_{AdS}$ is the $AdS_5$ radius, $l_s$ is a string length,
%$g_s$ and $g_{SYM}$ are a string and a SYM coupling constant, $N$ is the rank
%of the gauge group $SU(N)$ and $\lambda$ is the 't Hooft parameter.
%
% From the expressions \p{para} one can deduce different limits of 
%the parameters for which one or another approximation of string theory
%can be used for establishing the duality.
%
%When $R_{AdS}$ is fixed and  $\lambda\gg 1$, classical supergravity is 
%a good approximation, while when $\lambda$ is fixed and $N\rightarrow \infty$
%the classical superstring theory is a good approximation for the description
%of SYM theory. 
%%%%%%%%%%%%%%%%%%%%

For more profound analysis and check of this correspondence it is desirable
to know the detailed structure of the bulk theory 
(for instance, IIB superstring in $AdS_5\times S^5$) and the 
structure of the superconformal theory on the boundary.
This is especially essential for understanding field interactions.

This information  can be provided by 
corresponding effective actions invariant 
under superconformal transformations.
A natural way of getting such actions is 
to consider the dynamics of superbranes 
in $AdS\times S$ backgrounds, from which, in fact, the AdS/CFT conjecture 
originated. 

In this connection an extensive work has been undertaken by several 
theoretical  groups to derive superconformal field theory actions 
from $d=p+1$ worldvolume 
actions of superbranes in $AdS_{p+2}\times S^{D-p-2}$ backgrounds of 
D--dimensional supergravity \cite{2bs}--\cite{zhou1}. 

The idea is rather simple and natural. The 
$AdS_{p+2}\times S^{D-p-2}$ superbackground has isometry symmetry 
described by a 
supergroup $G$. For instance, in the case of an M--theory membrane, 
when $p=2$ 
and $D=11$, $G$ is an orthosimplectic supergroup $OSp(8|4)$; 
in the M5--brane 
case
$p=5, D=11$ and $G=OSp(2,6|4)$; and in the case of 
the IIB D3--brane $p=3$, $D=10$ 
and $G=SU(2,2|4)$. (The numbers in the names of the supergroups correspond to 
their bosonic subgroups). The maximal bosonic subgroup of $G$
(which is the bosonic isometry of $AdS_{p+2}\times S^{D-p-2}$) is 
$SO(2,p+1)\times SO(D-p-1)$, and the fermionic transformations of $G$ are 
generated by 32 Grassmann generators $\hat Q_{\alpha}$. Now recall that 
$SO(2,p+1)$ is also a group of conformal transformations acting on a 
(p+1)--dimensional space--time ${\cal M}_{p+1}$ 
associated with the boundary of 
$AdS_{p+2}$. In the approach under 
consideration the role of ${\cal M}_{p+1}$ is 
played by the $d=p+1$ worldvolume of the corresponding superbrane. The whole 
supergroup
$G$ will thus be associated with superconformal symmetry of a worldvolume
field theory describing the dynamics of the physical modes of the superbrane.

Consider in detail how the superconformal 
theory is derived from the superbrane 
action with the use of the example of an M2--brane in $AdS_4\times S^7$. 
All other more complicated 
cases (D3--brane, M5--brane etc.) can be treated in a similar way.

Let us start with the discussion of general properties of 
superbrane worldvolume 
actions. The action for a supermembrane propagating in a $D=11$ supergravity 
background has the following form \cite{bst}
\begin{equation}\label{2}
S_{M2}=-\int_{{\cal M}_3}d^3\xi\sqrt{-\det{g_{ij}}}-\int_{{\cal M}_3}d^3\xi
\varepsilon^{ijk}\partial_iZ^L\partial_jZ^M\partial_kZ^NA_{NML}(Z),
\end{equation}
where 
$g_{ij}(\xi)=\partial_iZ^ME_M^{  a}(Z)E_{{  a}N}(Z)\partial_jZ^N$ 
($i,j=0,1,2$; $  a=0,1,...,10$) 
is a worldvolume metric induced by embedding 
${\cal M}_3$
(parametrized by $\xi^i$) into a curved 
$D=11$ target superspace parametrized by 
bosonic coordinates $X^{  m}$ ($  m=0,1,...,10$)
and fermionic Majorana spinor coordinates $\Theta^{ \alpha}$
($  \alpha=0,1,...,32$) called all together $Z^M$.

$E^{  a}=dZ^ME^{  a}_M(Z)$ and $E^{  \alpha}
=dZ^ME^{ \alpha}_M(Z)$ 
are supervielbeins describing the geometry of the target superspace. Their 
leading components correspond to the graviton $e^{  a}_{  m}(X)=
E^{  a}_{  m}|_{\Theta=0}$ and the gravitino $\psi^{ \alpha}_{  
m}(X)=E^{ \alpha}_{  m}|_{\Theta=0}$, and $A_{NML}(Z)$ is a three-form
superfield whose leading component $A_{  n  m  l}(X)=A_{  n  
m  l}(Z)|_{\Theta=0}$ is the gauge field of $D=11$ supergravity. 

The action \p{2} is invariant under target--space superdiffeomorphisms
\be\label{3}
Z'^M=Z'^M(Z),
\ee
local worldvolume diffeomorphisms
\be\label{4}
\xi'^i=\xi'^i(\xi)
\ee
and local fermionic $\kappa$--symmetry transformations 
%having the following properties
\be\label{5}
\delta_\kappa Z^ME_M^{  a}=0, \quad 
\delta_\kappa Z^ME_M^{  \alpha}=\kappa^{ \beta}(\xi)
(1+\bar\Gamma)_{ \beta}^{~ \alpha},
\end{equation}
where
\be\label{61}
\bar\Gamma={1\over{6\sqrt{-g}}}\varepsilon^{ijk}
\Gamma_{ijk}, \qquad \bar\Gamma^2\equiv 1
\ee
and hence $1+\bar\Gamma$ is a spinor projection matrix. $\Gamma_{ijk}$ is
an antisymmetric product of $D=11$ gamma--matrices $(\Gamma_{a})^{~~\a}_{\b}$ 
pulled back on to the worldvolume, i.e 
$\Gamma_i\equiv\partial_iZ^ME_M^{  a}\Gamma_{  a}$. 

The appearance of the spinor projector in the 
$\kappa$--transformations reflects 
the fact that the presence of the supermembrane 
in the target superspace breaks 
half the 32 supersymmetries of a $D=11$ supergravity vacuum, the unbroken 
supersymmetries being associated with those Grassmann coordinates 
$\Theta^{ \a}$ which can be eliminated 
by $\kappa$--symmetry transformations, 
while remaining 16 $\Theta^{ \a}$ 
are worldvolume Goldstone fermions of the 
spontaneously broken supersymmetries and describe physical fermionic modes of 
supermembrane fluctuations.

%\bigskip
%{\bf Digression}
%
%To understand this symmetry breaking 
%in more detail consider an infinite flat membrane in a 
%$D=11$ Minkowski superspace. The presence of the membrane breaks
%
%ii) $D=11$ Lorenz symmetry down to Lorentz symmetry $SO(1,2)$ 
%of the worldvolume
%times $SO(8)$ rotations in directions orthogonal to the membrane;
%
%ii) $D=11$ Poincare invariance down to a translational symmetry along the 
%membrane worldvolume;
%
%iii) therefore, unbroken supersymmetries are those generated by supercharges 
%whose anticommutators close on translations along the membrane worldvolume. 
%Corresponding target superspace Grassmann coordinates can be regarded as ones
%parallel to the membrane.
%\bigskip
%
An important requirement for the $\kappa$--transformations \p{5} to be a 
symmetry of the membrane action \p{2} is that the 
target--space supervielbeins 
$E^{  a}(Z)$, $E^{ \a}(Z)$, superconnections $\Omega^{~~a}_{b}(Z)$ 
and the gauge superfield
$A^{(3)}$ satisfy $D=11$ supergravity constraints. The most essential 
constraints are the torsion constraint
\be\label{6}
T^{  a}=dE^{  a}+ E^{  b}\Omega^{~~a}_{b}
=iE^{  \a}\Gamma^{  a}_{  \a  \b}E^{  \b},
\ee
and the field--strength $F^{(4)}=dA^{(3)}$ constraint
\be\label{7}
F^{(4)}={i\over 2}E^{  a}E^{  b}E^{  \a}E^{  \b}
(\Gamma_{  a  b})_{  \a  \b}
+{1\over{4!}}E^{  a}E^{  b}E^{  c}E^{  d}
F_{  a  b  c  d}.
\ee
Other constraints are either conventional or can be obtained from \p{6} and 
\p{7} by considering their Bianchi identities.

The $D=11$ supergravity constraints are amount to supergravity equations of 
motion. Therefore, a supergravity background compatible with membrane 
$\kappa$--symmetry must be a solution of supergravity field equations.
When the gravitino field is zero the supergravity equations are
the Einstein equations for the $D=11$ curvature and Maxwell--like equations
for $F^{(4)}$
$$
R_{  m  n}-{1\over 2}g_{  m  n}R=
{1\over 3}(F_{  m  l_1  l_2  l_3}
F_{  n}^{~  l_1  l_2  l_3}-{1\over 8}g_{  m  n}F^2);
$$
\begin{equation}\label{d11eq}
D_{  p}F^{  p  l  m  n}=
{1\over{576}}\epsilon^{  l  m  n  l_1...  l_8}
F_{  l_1  l_2  l_3  l_4}F_{  l_5  l_6  l_7  l_8}.
\end{equation} 

The $AdS_4\times S^7$ is one of the solutions of \p{d11eq}
found almost twenty years ago
by Freund and Rubin \cite{fr} with the purpose to compactify $D=11$ 
supergravity ala Kaluza and Klein. For such a solution the gravitino 
field $\psi^{  \a}_{  m}(X)$ is zero.

It has been known that this solution is invariant under the maximum number of 
supersymmetry transformations  whose 32 parameters satisfy an 
$AdS_4\times S^7$ Killing spinor condition. 
%%%%%%%%%%%%%%%%%%%%%%%%%
%$\varepsilon^{ \a}$ 
%satisfy a Killing spinor condition
%\be\label{8}
%  D_{  m}\varepsilon=(\partial_{  m}
%-{1\over 4}\omega_{  m}^{  a  b}
%\Gamma_{  a  b}
%+T_{  m}^{~  n  p  q  r}F_{  n  p  
%q  r})\varepsilon(X)=0,
%\ee
%where 
%$$
%T_{  m}^{~  n  p  q  r}={1\over{288}}(\Gamma_{  m}^{~  
%n  p  q  r}
%-8\delta_{  m}^{[  n}\Gamma^{  p  q{  r}]}).
%$$
%and $\omega_{  m}^{  a  b}(X)$ 
%is the leading bosonic component of the 
%superconnection $\Omega_{  m}^{  a  b}(Z)$.
%
%The Killing spinor condition implies that the gravitino field 
%does not transform
%under these supersymmetry transformations 
%$ (\delta\psi^{ \a}_{  m}\equiv
%({  D}_{  m}\epsilon)^{ \a}=0)$ and remains zero. 
%Thus, together with a 
%flat Minkowski space--time, the $AdS_4\times S^7$ configuration has been 
%considered
%as a possible classical supersymmetric (Kaluza--Klein) vacuum of $D=11$ 
%supergravity.
%%%%%%%%%%%%%%%%%%%%%%%%%%%%%%%%%%%%

As a metric on $AdS_4\times S^7$ it is convenient to take the following one
\begin{equation}\label{9}
ds^2 = \left({r\over R}\right)^{4} dx^i \eta_{ij} dx^j 
+ \left(\frac Rr\right)^2 dr^2 + R^2 d\Omega^2\,,
\end{equation}
where $x^i,r$ $(i=0,1,2)$ are coordinates of the $AdS_4$ and $d\Omega^2$
stands for a  metric of the sphere $S^7$ of a radius $R$ parametrized by 
coordinates $y^{a'}$ $(a'=1',...,7')$. The coordinates $x^i$ of $AdS_4$
will be identified with the worldvolume coordinates $\xi^i$ upon imposing a
static gauge.

When $r\rightarrow\infty$ the second term in \p{9} tends to zero and
effectively the AdS part of the metric becomes three--dimensional. The
flat $d=3$ metric $dx^i \eta_{ij} dx^j$ is associated with the $AdS_4$ boundary
${\cal M}_3$ which is a Minkowski space.

For this choice of the $AdS_4\times S^7$ metric the gauge field 
$A^{(3)}(X)=A^{(3)}(Z)|_{\Theta=0}$ and its field strength $F^{(4)}$, which
satisfy the $D=11$ supergravity equations, have the following components which
are non--zero only in the $AdS_4$ part of space--time
\begin{equation}\label{10}
A^{(3)}=dx^2dx^1dx^0({r\over R})^{6}, \quad
F^{(4)}=-6drdx^2dx^1dx^0({{r^5}\over {R^{6}}})
\end{equation}

When the supermembrane propagates in the $AdS_4\times S^7$ supergravity 
background its action \p{2} is invariant under the supergroup $OSp(8|4)$ of the 
isometries of this background, to which the target--space superdiffeomorphisms 
\p{3} are reduced. A  superconformal form 
of the $osp(8|4)$ superalgebra will be presented a bit later.

At this stage $OSp(8|4)$ is not yet the superconformal group which acts on the 
worldvolume ${\cal M}_3$ of the membrane but it is rather an internal symmetry 
of the fields living on ${\cal M}_3$. For the  $OSp(8|4)$ supergroup to become
a worldvolume superconformal symmetry one should (in an appropriate way) 
gauge fix the local worldvolume diffeomorphisms \p{4} and the 
$\kappa$--symmetry \p{5} of the
membrane action. This gauge fixing eliminates the pure--gauge 
degrees of freedom
so that only fields which correspond to the physical modes of the superbrane 
remain in the theory.

An appropriate condition for fixing the worldvolume diffeomorphisms
is a static gauge when worldvolume coordinates are identified with three 
coordinates of $AdS_4$
\be\label{11}
\xi^i=x^i \quad(i=0,1,2)
\ee
(which at $r\rightarrow\infty$ parametrize the $AdS_4$ boundary). Thus the 
membrane worldvolume is associated with the $AdS_4$ boundary. 
In the static gauge eight physical bosonic worldvolume fields are 
the AdS radial coordinate $r(\xi)$ and the $S^7$ coordinates $y^{a'}(\xi)$.

A possible gauge fixing of $\kappa$--symmetry \p{5} compatible with the
static gauge \p{11} is putting to zero the following 16 components of 
$\Theta^{ \a}$
\be\label{12}
\eta^{ \a}\equiv[(1+\Gamma^{012})\Theta]^{ \a}=0.
\ee
Then other 16 coordinates
\be\label{121}
\theta^{ \a}\equiv[(1-\Gamma^{012})\Theta]^{ \a}
\ee
remain as the fermionic physical fields on ${\cal M}_3$. 
Note that in the gauge \p{11} $1+\Gamma^{012}$ coincides with the 
$\kappa$--symmetry projector
$1+\bar\Gamma$ when the membrane is in a static (vacuum) state,
i.e. when the bosonic fields $(r, y^i)$ 
are worldvolume constants and
$\theta^{ \a}$ are zero.
This makes the $\kappa$--symmetry gauge \p{12} admissible.

A combination of the worldvolume bosonic \p{4} and fermionic \p{5} 
transformations accompanied by the target superspace isometry transformations
of $OSp(8|4)$ which preserves 
the gauge fixing conditions \p{11} and \p{12} now becomes
a non--linearly realized superconformal symmetry on the supermembrane 
worldvolume. This is because we have identified the worldvolume coordinates
with target space AdS coordinates, and the parameters of the worldvolume 
symmetries are now expressed in terms of the constant parameters of target 
superspace symmetry $OSp(8|4)$. For instance, the static gauge \p{11} is
preserved if the parameters of the worldvolume transformations 
$\delta_{w}$ and $\kappa$-transformations $\delta_{\kappa}$ are 
related to the parameters of $OSp(8|4)$ transformations $\delta_{osp}$ as 
follows
$$
0=x^i-\xi^i=x'^i-\xi'^i=x^i-\xi^i +\delta_{osp}x^i+\delta_w x^i+
\delta_\kappa x^i-\delta_w\xi^i
\quad \Rightarrow 
$$
\be\label{13}
\quad \delta_w x^i-\delta_w\xi^i+\delta_\kappa x^i
=- \delta_{osp}x^i.
\ee

To find an explicit form of the superconformal variations of the 
physical fields
$r(\xi)$, $y^{a'}(\xi)$ and $\theta^{ \a}(\xi)$ and to derive from the 
original
superbrane action the superconformal action describing the dynamics of these 
fields one should know an explicit form of the supervielbeins 
$E^{  a}(Z)$, $E^{ \a}(Z)$ and of
the three--form superfield $A^{(3)}(Z)$ which enter the supermembrane action.
This is a key point in the construction of the superconformal action.

At the moment we know only the leading components of these objects at 
$\Theta=0$. These are the bosonic $AdS\times S$ metric \p{9} and the bosonic 
value
of the $A^{(3)}(x,r)$ field \p{10}. 

A direct way of getting $E^{  a}(Z)$, $E^{ \a}(Z)$ and $A^{(3)}(Z)$ 
is to solve the supergravity
constraints \p{6} and \p{7} taking the values \p{9} and \p{10} of the 
superforms 
at $\Theta=0$ as initial conditions. This has been done by Claus
\cite{claus}. However, this explicit form
is rather complicated since $E^{  a}(X,\Theta)$ and $E^{ \a}(X,\Theta)$ 
are polynomials of up to the 32-nd
power in $\Theta$. Even if half of $\Theta$ are eliminated by gauge fixing 
$\kappa$--symmetry, this will, in general, result in polynomials of the 16-th 
power. So, if written in terms of these polynomials the supermembrane action
looks very cumbersome and untreatable. Hence it is desirable to find simpler 
form of the AdS superforms. 

An elegant way of looking for this form is to consider the $AdS_4\times S^7$ 
superspace as a so called coset superspace.

The method of coset spaces has been applied to the description 
of $AdS\times S$ 
superspaces in a number of papers \cite{2bs,2bd3,mbra,pst4,zhou1} 
and is based on a classical work of 
Cartan on group manifolds, symmetric and homogeneous spaces. It is worth 
mentioning that the first physical model with global supersymmetry \cite{va} 
and the first supergravity \cite{vs} was constructed by using this method.

Let us briefly sketch basic ideas of this method by the use of the example of 
the AdS superspace. As we have already discussed the $AdS_4\times S^7$ 
superspace has the isometry symmetry described by the supergroup $OSp(8|4)$. 
This means that a given point $Z^M=(X^{  m},\Theta^{ \a})$ of this 
superspace can be connected with another point by 
an $OSp(8|4)$ transformation. 
A subgroup of $G=OSp(8|4)$ which leaves 
a given point $Z^M$ intact is called the 
stability (or isotropy) subgroup of $G$. In our case the stability group is a 
bosonic group $H=SO(1,3)\times SO(7)$ which acts on a superspace tangent to 
$AdS_4\times S^7$. The connection forms which 
define the parallel transport in the 
$AdS$ superspace take their values in the algebra of $H$.

The space with these properties is called the coset space and is
denoted as
$$
K=G/H.
$$
It is the space of the classes of equivalence of the points of $G$ which
are related by the $H$--transformations.

In our case 
$$
K={{OSp(8|4)}\over{SO(1,3)\times SO(7)}}.
$$
Note that the bosonic subspaces $AdS_4$ and $S^7$ of this superspace are coset
spaces themselves, actually, they are symmetric spaces
$$
AdS_4={{SO(2,3)}\over{SO(1,3)}}, \qquad S^7={{SO(8)}\over{SO(7)}}.
$$
Note also that the difference between the dimensions of 
the group $G$ and $H$ is 
equal to the (bosonic plus fermionic) dimension of the coset space
$$
dim~ K= dim ~G-dim~H.
$$
Thus the generators of $G$, called coset generators $K$, 
which are not contained 
in the stability subgroup of $H$, 
are in one to one correspondence with the coset
space coordinates $Z^M$ and are associated with the boosts 
of the points of the coset superspace.

Now remember that an element of the group $G$ can be 
represented as an exponent 
of the generators $G_I$ of the group times parameters of the group 
transformations $\lambda^I$, considered as coordinates of the group manifold 
$G$
\be\label{140}
G(\lambda)=\exp{(\lambda^IG_I)}.
\ee
By analogy, we can identify a point $Z^M=(X^{  m},\Theta^{ \a})$ of the 
AdS superspace with a coset element $K(Z)$ realized as an exponent
\be\label{14}
K(Z)=\exp{(X^{  a}P_{  a}+\Theta^{ \a}{\hat Q}_{ \a})},
\ee
where $P_{a}=(P_i, P_r;P_{a'})$ $(i=0,1,2$, $a'=1,...,7$) 
are bosonic boosts 
acting, respectively, on $AdS_4$ and $S^7$, and $  Q_{ \a}$ are 
supertranslations (supercharges). 

To get the variation properties of the $AdS$ superspace coordinates $X^{  
m},\Theta^{ \a}$ under $OSp(8|4)$ one should act on $K(Z)$ by a supergroup 
element $G(\lambda)$
$$
K'(Z'(\lambda,Z))=G(\lambda)K(Z).
$$

If we now use $K(Z)$ to construct a so called Cartan one--form
$
L=K^{-1}(Z)dK(Z),
$
it is claimed that this form takes values in the algebra of $G$, and its 
components are supervielbeins and superconnections describing the geometry of
the AdS superspace
\be\label{15}
L=K^{-1}(Z)dK(Z)=E^{  a}P_{  a}+E^{ \alpha}{\hat Q}_{ \alpha}
+\Omega^{  a  b}M_{  a  b}.
\ee
The one--forms $E^{a},E^{ \alpha}$ corresponding to the boost
generators $P_{a}, {\hat Q}_{ \alpha}$ are supervielbeins (which we
can use to construct the supermembrane action), while the one--forms 
$\Omega^{  a  b}$ are connections in the $AdS$ superspace taking their 
values in the stability subalgebra $h=so(1,3)\times so(7)$ 
generated by $M_{  
a  b}$. Note that the indices $  a$ and $ \a$ correspond now to the 
vector and a spinor representation of the stability group, respectively.

To convince oneself that $E^{  a},E^{ \alpha}$ and 
$\Omega^{  a  b}$ 
have indeed the properties of vielbeins and connections, consider their 
transformation properties under the action of the stability group H, 
which acts 
on $K(Z)$ from the right:
$$
K'(Z)=K(Z)H(Z),
$$
\be\label{16}
L'=K'^{-1}dK'=H^{-1}(K^{-1}dK)H+H^{-1}dH.
\ee
Comparing \p{15} with \p{16} we see that $E^{  a},E^{ \alpha}$ 
transform homogeneously under
H--transfor\-mations, while $\Omega^{  a  b}$ acquires an inhomogeneous 
contribution from $H^{-1}dH$ which takes values in the algebra of the 
generators $M_{  a  b}$. Thus, $E^{  a},E^{ \alpha}$ and 
$\Omega^{  a  b}$ indeed have the correct transformation properties of 
supervielbeins and superconnections under the 
tangent superspace stability group.

Now, taking an exponential parametrization \p{14} of the coset element $K(Z)$
(which is in fact not unique and corresponds to a choice of a 
local supercoordinate 
system), substituting it into the expression for the Cartan form \p{15} and 
making use of the exact form of the superalgebra of the
$OSp{(8|4)}$ generators one can derive an explicit form of the supervielbeins 
and superconnections. In general they will be again polynomials of the 32-nd 
power in $\Theta$, as has been obtained by Metsaev and Tseytlin \cite{2bs}. 
And further work is 
required to check whether this polynomial dependence can be reduced down to 
lower powers in $\Theta$ due to some matrix identities, which is 
{\it a priori} not obvious.

An alternative way is to find a suitable exponential parametrization of the 
coset element $K(Z)$ which would directly produce the supervielbeins and 
superconnections as short polynomials in $\Theta$. Such a 
parametrization has been found in \cite{pst4}.
The idea has been to arrange the generators of $OSp(8|4)$ in such a way that 
their (anti) commutators take the explicit form of the superconformal 
algebra acting on a 
$d=3$ subspace ${\cal M}_3$ of $AdS_4$. So let us take the following 
generators as the coset superspace 
generators which appear in the exponent of $K(Z)$ 

$\Pi_i$ $(i=0,1,2)$ -- the boost (momenta) generators on $M_3$;

$P_r=D$ -- the dilatation generator;

$P_{a'}$ $(a'=1,...,7)$ -- boosts on the $S^7$ sphere;

$Q_{ \a}=(1+\Gamma^{012}) \hat Q_{ \a}$ -- ordinary supersymmetry 
transformations;

$S_{ \a}=(1-\Gamma^{012})\hat Q_{ \a}$ -- special superconformal 
transformations.

To close the $OSp(8|4)$ superalgebra we must add to these generators the 
generators which correspond to the stability group $SO(1,3)\times SO(7)$.
These consist of $SO(7)$ rotations $M_{a'b'}$, $SO(1,2)$ rotations $M_{ij}$ 
plus boost generators $M_{ir}=\Pi_i-K_i$ which
all together form the generators $M_{ab}$ of the $SO(1,3)$--rotations, and 
the generators $K_i$ of special conformal transformations of ${\cal M}_3$.

The $osp(8|4)$ superalgebra written in terms of these generators has a 
relatively simple form
\be\label{17}
[\Pi_i,\Pi_j]=0,~ [\Pi_i,Q_{ \a}]=0,~ \{Q,Q\}\sim \Gamma^i\Pi_i,
~[D,\Pi_i]=\Pi_i,~ \{D,Q_{ \a}\}={1\over 2}Q_{ \a},
\ee
\be\label{18}
[K_i,K_j]=0,\quad [K_i,S_{ \a}]=0, \quad \{S,S\}\sim -\Gamma^iK_i,
\quad [D,K_i]=-K_i, \quad 
\ee
$$
\{D,S_{ \a}\}=-{1\over 2}S_{ \a},
$$
$$
[\Pi_i,S]\sim - \Gamma_iQ, \quad [K_i,Q]\sim \Gamma_iS,
$$
\be\label{19}
\{Q_{ \a},S_{ \b}\}=h^{~~~A}_{ \a \b}T_A, \quad
[T_A,Q_{ \a}]=t_{A \a}^{~~~ \b}Q_{ \b}, \quad
[T_A,S_{ \a}]=g_{A \a}^{~~~ \b}S_{ \b},
\ee
where $T_A$ stand for the generators $D$, $M_{ij}$, $P_{a'}$ and $M_{a' b'}$.
To close the $osp(8|4)$ superalgebra one should add to \p{17}--\p{19} 
commutation relations of $\Pi_i$ and $K_i$ with $T_A$. But it is not 
necessary 
to know these commutators explicitly for the derivation of the Cartan 
superform.

We see that the commutation relations of $\Pi_i$ and $Q$, 
and $K_i$ and $S$ form 
three--dimensional super Poincare subalgebras of $osp(8|4).$

Now, as the exponent parametrization of the coset $K(Z)$ let us take
\be\label{20}
K=e^{{x^i}\Pi_i}e^{(\log{r\over R})D}e^{y^{a'}P_{a'}}
e^{\eta^{ \alpha}Q_{ \alpha}}e^{\theta^{ \alpha}S_{ \alpha}},
\end{equation}
where $\eta=(1+\Gamma^{012})\Theta$ and $\theta=(1-\Gamma^{012})\Theta$ are 
projected  $AdS$ Grassmann coordinates already considered above in connection 
with $\kappa$--symmetry gauge fixing \p{12}.

Using this parametrization and the form \p{17}--\p{19} of the $osp(8|4)$ 
commutation relations we can compute the components of the Cartan form
$K^{-1}dK$ which appear to be polynomials of only 6-th power in $\theta$ and 
$\eta$. We can simplify the form of the supervielbeins even more 
if we gauge fix 
the $\kappa$--symmetry of the supermembrane in a suitable way. As we have 
already discussed (eq. \p{12}) a possible gauge choice is to put to zero 
the supercoordinates $\eta$ (eq. \p{12}). After such a gauge fixing the 
$\eta$--exponent drops out of $K(Z)$ in \p{20} and the resulting Cartan form 
components
become polynomials of the 4-th power in $\theta$.

For instance, the vector supervielbeins which form the induced worldvolume 
metric have the following structure (for simplicity we skip numerical 
coefficients)
\be\label{21}
E^i(x,r,y,\theta)=\left({r\over R}\right)^2 dx^i+iD\theta\Gamma^i\theta,
\ee
$$
E^r={R\over r}dr + \left({r\over R}\right)^2 dx^i\theta\Gamma_ih^{r}\theta,
$$
$$
E^{a'}(x,r,y,\theta)=e^{a'}_{S^7}(y)+\left({r\over R}\right)^2 
dx^i\theta\Gamma_ih^{a'}\theta,
$$
$$
D\theta^{ \a}=d\theta^{ \a}+E^A(\theta g_A)^{ \a}+\left({r\over 
R}\right)^2 dx^i(\theta\Gamma_ih^{A}\theta)(\theta g_A)^{ \a},
$$
where $E^A=e^A(x,r,y)+\left({r\over R}\right)^2 
dx^i(\theta\Gamma_ih^{A}\theta)$ 
are supervielbeins and superconnections
associated with the generators $D$, $P_{a'}$ and $M_{  a  b}$ and 
$e^A(x,r,y)$ are their bosonic values at $\theta=0$.

%%%%%%%%%%%%%%%%%%%
%The supervielbein spinor components are
%$$
%[(1+\Gamma^{012})E]^{ \a}=\left({r\over R}\right)^2 
%dx^i(\theta\Gamma_i)^{ \a},
%$$
%\be\label{211}
%[(1-\Gamma^{012})E]^{ \a}=D\theta^{ \a}+\left({r\over R}\right)^2 
%dx^i(\theta\Gamma_ih^{A}\theta)(\theta g_A)^{ \a}.
%\ee
%%%%%%%%%%%%%%%%%%%%%

We see that supervielbeins and superconnections have 
a relatively simple form, 
though if we substitute them into the supermembrane action we will 
get an action 
for a superconformal field theory which still has a rather 
complicated structure
of field interactions, which hinders the analysis of this theory.

Much simpler form (up to the second power in $\Theta$) 
of the supervielbeins and 
superconnections might be obtained if in the exponent representation \p{20}, 
instead of putting to zero $\eta$ we might put to zero $\theta$.
Such a $\kappa$--symmetry gauge choice would be compatible with an 
(anti)static
gauge of the worldvolume diffeomorphisms, when, for example, the worldvolume
time is identified with minus $AdS$ time coordinate
\be\label{22}
\xi^0=-x^0, \quad \xi^1=x^1, \quad \xi^2=x^2.
\ee
Then we would have
\be\label{23}
E^i=({r\over R})^2(dx^i+id\eta\Gamma^i\eta), \quad
E^r=e^r(x,r,y), \quad
E^{ \a}={r\over R}d\eta^{ \a},
\ee
which look very much like flat superspace covariant superforms.

This simple $\kappa$--symmetry gauge was proposed in \cite{mbra} and \cite{k}, 
and is called a supersolvable algebra 
gauge, or a Killing spinor gauge. It is called supersolvable since when 
$\theta=0$ the remaining generators in the parametrization of 
the coset element
$K(Z)$ in \p{20} form a sub--superalgebra of $osp(8|4)$, which is an extension
of a $d=3$ super--Poincare algebra by the dilatation generator $D$.

And it is called a Killing spinor gauge since it corresponds to an appropriate
choice of the solution of the AdS Killing spinor equation. 
%\p{8}.

In this gauge the supermembrane action would take  much simpler form, 
but here
appears a problem that this gauge is not always admissible. 
For instance, it is
not compatible with a natural static vacuum solution of the superbrane 
equations of motion when
\be\label{24}
\xi^i=x^i, \quad r=const,\quad y^{a'}=const, \quad \Theta=0,
\ee
i.e. when the brane completely lives in a three--dimensional slice of the 
$AdS_4$ space--time. For such a solution $\eta=0$ is an admissible gauge.
Note that the static vacuum solution \p{24} is a BPS saturated state since 
it is invariant under the 16 standard
supersymmetries $Q_{ \a}$ and (spontaneously) 
breaks special superconformal 
symmetry $S_{\a}$, which is nonlinearly realized on the excitations 
over this vacuum solution. 

The $\theta=0$ gauge would be compatible with an (anti) static configuration 
with the reverse orientation of time (or a space) coordinate, i.e.
\be\label{25}
\xi^0=-x^0, \quad \xi^1=x^1, \quad \xi^2=x^2, 
\quad r=const,\quad y^{a'}=const, 
\quad \Theta=0,
\ee
but this is not a solution of the superbrane equations, unless $r$ is zero. 
And 
when $r=0$ such a brane configuration shrinks to a point at the $AdS$ horizon.
Physically this configuration describes an antibrane which is attracted by
a bunch of branes whose metric near horizon is close to that of $AdS\times S$.

There may exist, however, anti--static brane configurations, 
compatible with the
Killing gauge, which extend along the radial coordinate of $AdS$ and/or 
somehow 
nontrivially wind around compact directions of the sphere. In general, such 
brane configurations will break all supersymmetries of the superbackground. 
However, an action which describes fluctuations over these configurations 
will have a simple
fermionic structure due to the simple form of the supervielbeins \p{23}, 
and it would be of interest to study the properties of such theories.

The above examples teach us that the gauge choice is a subtle point 
and depends on which classical solution of field equations one deals with.
Recently this problem has been also discussed in the case of a D0--brane
in $AdS_2\times S_2$ \cite{zhou2}.

An example of the use of the Killing spinor gauge $(\theta=0$)  is a IIB 
superstring propagating in the $AdS_5\times S^5$ superbackground. 
The $AdS_5\times S^5$ superbackground can be viewed as a large N limit of
coincident D3--branes. The Killing gauge is compatible with superstring
$\kappa$--symmetry since the $\kappa$-symmetry projector now differs from
the one used to impose the Killing gauge, the latter being related to
the D3--brane $\kappa$--symmetry projector.

This theory has been considered by Pesando, Kallosh and Rahmfeld,  and 
Kallosh and Tseytlin \cite{2bs}. 

Using the $AdS_5\times S^5$ supervielbeins in the Killing gauge, which have the
 form analogous to that written in \p{23}, one can obtain the following 
superconformal action for the IIB superstring
\begin{eqnarray}
S =-\frac{1}{2}\int d^2\xi\ \biggl[\sqrt{-g} \, g^{ij}&& \hspace{-0.7cm}
\left(y^2(\partial_i x^p
- 2 i \bar
\eta \Gamma^{p} \partial_i\eta)(\partial_j x^p - 2 i \bar
\eta \Gamma^{p} \partial_j \eta) +\frac{1}{y^2} \partial_i y^t
\partial_j y^t \right) \nonumber \\ &&
 +\  4 i \eps^{ij} \partial_i y^t \bar \eta \Gamma^t
\partial_j\eta \ \biggr] \ , 
\label{SA}
\end{eqnarray} 
where now $\xi^i$ $(i=0,1)$ parametrize the superstring worldvolume,
$g_{ij}(\xi)$ is an intrinsic (auxiliary) worldvolume metric, $x^p$ 
$(p=0,1,2,3)$ denote coordinates parallel to the D3--branes; 
$y^t=(r,y^{a'})$, 
which include the AdS radial coordinate and the $S^5$ coordinates, 
stand for the 
coordinates orthogonal to the D3--branes, and $y^2=y^ty^t$.

This action differs from a IIB string action in a flat $D=10$ superbackground,
with $\kappa$--symmetry being gauge fixed in the same way as in \p{SA}, by the
factors $y^2$ and $1\over y^2$.

Now, having the action for a IIB superstring in $AdS_5\times S^5$ one can
study classical and quantum properties of this theory.
First steps in this direction were undertaken by Kallosh and Tseytlin.
It has been realized that it is not obvious that the superstring equations 
yielded by such an action admit supersymmetric classical solutions. 
%Thus the resulting theory may be non--supersymmetric.
It would be of interest to analyze
a duality relation of this theory with a $D=4$ Yang--Mills theory on 
the boundary of AdS. For this one should also know a corresponding 
D3--brane action in  
$AdS_5\times S^5$ which would produce the Yang--Mills theory. This action is 
still under construction due to problems discussed above.

In conclusion what we have learned about the superbranes in the 
AdS backgrounds 
is that different gauge choices for fixing $\kappa$--symmetry of the original 
brane action may result in different  worldvolume actions (and field 
theories). This is because we deal with topologically nontrivial backgrounds
such as $AdS\times S$, and the most gauge fixing conditions are only locally 
admissible, and/or implicitly reflect how the brane is embedded into the
background. The problem of finding globally defined conditions for gauge fixing
brane actions in $AdS\times S$ has been considered in \cite{rr}.
The light-cone  gauge formalism for theories in AdS spaces has been developed
in \cite{met}.

An unexpected observation which we have made is 
that actions which admit classical
vacuum configurations preserving effective worldvolume supersymmetry
have more complicated structure of the fermionic sector than actions
for which the existence of supersymmetric brane configurations is problematic.
%More work should be done to clarify this situation.

\medskip
{\bf Acknowledgements}. This work was partially supported by the 
European Commission TMR Programme ERBFMPX-CT96-0045 to which the authors 
are associated, and by INTAS Grant 96-308. D.S. acknowledges the financial 
support from the Alexander von Humboldt Foundation.


\begin{thebibliography}{99}
\bibitem{ads}
 J. Maldacena, {\em Adv. Theor. Math. Phys.\/} {\bf 2} 
(1998) 231;\\
S. S~Gubser, I. R.~Klebanov and A. M.~Polyakov, {\em Phys. 
Lett.\/} {\bf B428} (1998) 105;\\
E.~Witten, {\em Adv. Theor. Math. Phys.\/} {\bf 2} (1998) 253.\\
O. Aharony, S. S. Gubser, J. Maldacena, H. Ooguri, Y. Oz,
 Large N Field Theories, String Theory and Gravity, hep-th/9905111.
\bibitem{2bs}
 R. R.~Metsaev and A. A.~Tseytlin, {\em Nucl. Phys.\/} 
{\bf B533} (1998) 109;\\
I. Pesando, JHEP {\bf 9811} (1998) 002; {\em Mod. Phys. Lett.} 
{\bf A14} (1999) 343; JHEP {\bf 9902} (1999) 007;\\
R.~Kallosh and J.~Rahmfeld, {\em Phys. Lett.\/} {\bf B443} (1998) 143;\\
 R.~Kallosh and A. A.~Tseytlin, JHEP {\bf 9810} (1998) 016.
I. Oda, {\em Phys. Lett.\/} {\bf B444} (1998) 127; JHEP {\bf 9810} (1998) 015.
\bibitem{2bd3}
 R. R.~Metsaev and A. A.~Tseytlin, {\em Phys. Lett.\/} {\bf B436} (1998) 281;
\bibitem{mbra}
 G.~Dall'Agata, D.~Fabbri, C.~Fraser, P.~Fr\'e, P.~Termonia 
and M.~Trigiante, {\em Nucl. Phys.\/} {\bf B542} (1999) 157.
\bibitem{claus}
P. Claus, {\em Phys. Rev.\/} {\bf D59} (1999) 066003.
\bibitem{pst4}
 P. Pasti, D. Sorokin and M. Tonin,
{\em Phys. Lett.\/} {\bf B447} (1999) 251.
\bibitem{zhou1}
Jian-Ge Zhou, Super 0-brane and GS Superstring Actions on $AdS_2 \times S^2$,
hep--th/9906013.
\bibitem{bst}
E. Bergshoeff, E. Sezgin and P. K. Townsend, 
{\em  Phys. Lett.\/} {\bf 189B} (1987) 75; {\em Ann. Phys.\/} {\bf 185} 
(1988) 330.
\bibitem{fr}
P. G. O. Freund and  M. A. Rubin, {\em Phys. Lett.\/} {\bf B97} (1980) 233.
\bibitem{va} 
D. V. Volkov and V. P. Akulov, {\em JETP Letters} {\bf 16} (1972) 438;
{\em Phys. Lett. \/} {\bf B46} (1973) 109.
\bibitem{vs}
D. V. Volkov and V. A. Soroka, {\em JETP Letters} {\bf 18} (1973) 312.
\bibitem{k}
R. Kallosh, Superconformal Actions in Killing Gauge, hep-th/9807206.
\bibitem{zhou2}
 M. Kreuzer and Jian-Ge Zhou, 
Killing gauge for the 0-brane on $AdS_2 \times S^2$ coset superspace,
hep-th/9910067.
\bibitem{rr} 
A. Rajaraman and  M. Rozali,  
On the Quantization of the GS String on $AdS_5 \times S^5$, hep-th/9902046.
\bibitem{met}
R. R. Metsaev, Light cone form of field dynamics in anti-de Sitter 
spacetime and AdS/CFT correspondence,  hep-th/9906217.
\end{thebibliography}
\end{document}